\documentclass[twocolumn]{aastex631}

\newcommand{\kms}{$\rm km~s^{-1}$} 
\newcommand{\lya}{Ly$\alpha$}

\newcommand{\HI}{\mbox{H\,{\sc i}}}

\newcommand{\OI}{\mbox{O\,{\sc i}}}
\newcommand{\OII}{\mbox{O\,{\sc ii}}}

\newcommand{\OVI}{\mbox{O\,{\sc vi}}}

\newcommand{\SiIII}{\mbox{Si\,{\sc iii}}}

\newcommand{\MgII}{\mbox{Mg\,{\sc ii}}}

\newcommand{\logm}{${\rm log}_{10}(M_{\star}/\rm M_{\odot})$}
\newcommand{\Msun}{$\rm M_{\odot}$}

\usepackage[flushleft]{threeparttable}
\usepackage{ragged2e}
\usepackage{graphicx}	% Including figure files
\usepackage{amsmath}	% Advanced maths commands
\usepackage{amssymb}	% Extra maths symbols
\usepackage{tabularx}
\usepackage{newtxtext,newtxmath,ulem}
\usepackage{multirow}
\usepackage{cases}
\usepackage{booktabs}%,caption}
\usepackage[T1]{fontenc}
\usepackage{xcolor}
\usepackage{soul}

\usepackage{dcolumn}
    \newcolumntype{d}[1]{D{.}{.}{#1}}

\begin{document}

\title{MUSEQuBES: The kinematics of \OVI-bearing gas in and around low-redshift galaxies}

\author[0009-0000-0797-7365]{Sayak Dutta}
\affiliation{Inter-University Centre for Astronomy \& Astrophysics, Post Bag 04, Pune, India 411007}

\author[0000-0003-3938-8762]{Sowgat Muzahid}
\affiliation{Inter-University Centre for Astronomy \& Astrophysics, Post Bag 04, Pune, India 411007}

\author{Joop Schaye}
\affiliation{Leiden Observatory, Leiden University, PO Box 9513, NL-2300 RA Leiden, the Netherlands}

\author{Sean Johnson}
\affiliation{Department of Astronomy, University of Michigan, 1085 S. University Ave, Ann Arbor, MI 48109, USA}

\author{Hsiao-Wen Chen} 
\affiliation{Department of Astronomy \& Astrophysics, The University of Chicago, 5640 S. Ellis Avenue, Chicago, IL 60637, USA}

\author{Sebastiano Cantalupo}
\affiliation{Department of Physics, University of Milan Bicocca, Piazza della Scienza 3, 20126, Milano, Italy}

\begin{abstract}

\noindent 
We present a detailed study of the kinematics of \OVI-bearing gas around 60 low-mass (median \logm~$=8.9$) galaxies at low redshift $(0.1 < z < 0.7)$ using background quasars (median impact parameter $\approx115$ kpc) as part of the MUSE Quasar-fields Blind Emitters Survey (MUSEQuBES). 
We find that the majority of the \OVI\ absorbers detected within the virial radius have line-of-sight velocities smaller than the escape velocities and are thus consistent with being gravitationally bound, irrespective of the halo mass. 
However, the fraction of such absorbers declines at larger impact parameters.
The Doppler $b$ parameter and the velocity width ($\Delta v_{90}$) of the \OVI\ absorbers exhibit large scatter inside the virial radius of the host galaxies, but the scatter declines sharply at impact parameter $D \gtrsim 2R_{\rm vir}$. 
For high-mass galaxies (\logm$>9$), \OVI\ absorption displays a larger kinematic spread, quantified by the pixel-velocity two-point correlation function (TPCF). However, this difference disappears for the isolated galaxies when the pixel velocities are scaled by the galaxy's circular velocity. 
We do not find any significant difference between the TPCF of isolated and group galaxies when the stellar mass is controlled for. 
A significant fraction of groups (4/6) with four or more member galaxies do not show any detectable \OVI\ absorption, likely due to the passive nature of nearest galaxies. 

\end{abstract}


\section{Introduction} \label{sec:intro}

The circumgalactic medium (CGM) serves as a vital reservoir of gas, influencing the life-cycle of galaxies through complex interactions and exchanges of matter and energy \citep[see][for a review]{Tumlinson_2017}. Due to the challenges of studying the diffuse CGM in emission, absorption line spectroscopy using bright background sources such as quasars has become the preferred method \citep[]{Bergeron_1986, Petitjean_1990}. The kinematics of the absorbing gas, as imprinted in the spectra of background quasars, provides crucial information on gas flows in and around galaxies. For example, absorbers with near solar metallicity exhibiting a large kinematic spread are thought to trace outflowing materials \citep[e.g.,][]{Tripp2011,Muzahid2015,Rosenwasser2018}. From the sign of the line of sight (LOS) velocity of absorption at the position of the absorber, it has been shown that the cool circumgalactic gas co-rotates with the galaxy, potentially representing accreting gas that is fueling the star formation \citep[e.g.,][]{Kacprzak2010,Bouche2013,Ho2017,Zabl2019}.  
%%% 

The fifth ionized state of oxygen (O$^{5+}$ or \OVI) is a favored diagnostic for studying the hotter phase of the CGM due to its doublet nature, large oscillator strengths, and the large cosmic abundance of oxygen. Although the peak ion-fraction ($f_{\OVI}$) in collisional ionization equilibrium (CIE) at $T\approx10^{5.5}$K makes \OVI\ an ideal tracer for warm-hot gas, it can also trace cooler, photoionized gas in photoionization equilibrium (PIE) with ionization parameter ${\rm log}_{10}~U\approx-1$. In addition, non-equilibrium processes can give rise to \OVI\ at lower temperatures for a medium with a sufficiently short cooling time  \citep[see][]{Oppenheimer_2013}.

The kinematics of \OVI\ absorption can provide crucial information regarding its origin and its connection to host galaxies \citep[but see][]{Kacprzak_2019}. Comparing the velocity centroids and widths of the \OVI\ absorbers with the inferred halo escape velocity of host galaxies is essential for understanding their origin(s) and the processes that drive metals far away from galaxies. Additionally, the Doppler$-b$ parameters of \OVI\ absorption components are crucial to constrain the temperature of the gas phase giving rise to the \OVI. Extensive investigations have explored the connection between bright galaxies and \OVI\ absorption using statistically significant samples of galaxy-absorber pairs \citep[e.g.,][]{Prochaska_2011,Tumlinson_2011,Turner_14,Kacprzak_2015,Johnson_15,Keeney_2017,Mishra_2024}.  The \OVI\ abundance and kinematics around foreground $\approx L_*$ galaxies in close projection to QSO sightlines is studied in COS-Halos survey \citep[]{Tumlinson_2011,Werk_2016}. They observed that the majority of their star-forming, but not passive galaxies exhibit \OVI\ absorption in their CGM. This dichotomy of \OVI\ absorption between star-forming and passive galaxies has since been reported even for stellar and halo mass-controlled samples in \citet[]{Tchernyshyov_2023}. The vast majority of the \OVI\ absorbers in the COS-Halos sample are found to be consistent with being bound to the host halos of $L_*$ galaxies with line-of-sight velocities considerably lower than the local escape velocities. \citet[]{Mathes_2014} and later, \citet[]{Ng_2019} found that \OVI\ absorbers are observed to be less bound from lower-mass galaxies than from high-mass halos.

The virial temperature of the host halos of $\approx L_*$ galaxies is ideal for producing \OVI\ in CIE. In contrast, the halos of lower mass (sub--$L_*$) galaxies are too cool for collisionally ionized \OVI.   Cosmological simulations have predicted that strong feedback processes, such as those from star formation and active galactic nuclei (AGN), can generate significant amounts of \OVI\ in the CGM \citep[e.g.,][]{Liang_2016}. The shallower potential wells of lower-mass galaxies make them more susceptible to these feedback processes. The kinematics of circumgalactic \OVI\ absorption carries imprints of such feedback processes. However, a comprehensive statistical sample of \OVI\ absorption around low-mass galaxies is still lacking in the literature. Recently, state-of-the-art integral field spectrographs (IFS) like MUSE \citep[]{Bacon_2010} on VLT/UT4 and KCWI \citep[]{Morrissey_2018} on Keck have paved the way for efficiently identifying low-mass galaxies and characterizing their environments around bright background quasars. As a consequence, the availability of \OVI\ measurements around galaxies with a large dynamic range of stellar mass and environment is gradually increasing \citep[see e.g.,][]{Qu_2024,Mishra_2024}.

\citet[]{Stocke_2014} and \citet{Stocke_2017} found that \OVI\ absorbers associated with groups require modeling with a few broad components, despite having an overall narrow absorption profile, compared to those associated with isolated galaxies. They attributed the broadness of the components to the warmer environment. However, the overall narrow absorption profile was explained by the inability to distribute \OVI\ over the ``circum-group'' medium, especially due to the instability of the \OVI\ phase caused by rapid cooling. They speculated that the inability to maintain a diffuse halo of warm CGM gas gives rise to a possible interface of the diffuse hot ($T > 10^6$~K) and cooler, photoionized regions embedded in the ``circum-group'' medium. Analysis based on the \OVI\ velocity two-point correlation function (TPCF) resulted in a similar conclusion \citep[see e.g.,][]{Pointon_2017, Ng_2019}. Since the overall kinematics of the CGM is expected to be strongly influenced by dynamical processes governed by gravity, it is crucial to study CGM absorption kinematics using stellar mass-controlled samples.

In this paper, we investigate the kinematics of \OVI-bearing gas around a unique, primarily low-mass sample of galaxies drawn from the low-$z$ part of the MUSEQuBES survey\footnote{The high-$z$ counterpart of MUSEQuBES studies the gas and metals around $z\approx3.3$ \lya\ emitters \citep[]{Muzahid_2021,Banerjee_2023}.} \citep[]{Dutta_2024}.  The relationships between host galaxy properties and column density, covering fraction, and mass of \OVI-bearing gas are presented in the companion paper Dutta et. al. (2024, submitted) [Paper I hereafter].
In Paper I, leveraging the detection and upper limits of \OVI\ column density ($N(\OVI)$), we showed that the average \OVI\ column density, ${\rm log}_{10}\left<N(\OVI)/{\rm cm}^{-2}\right> = 14.14^{+0.09}_{-0.10}$, within the virial radius of our low-mass MUSEQuBES sample is significantly lower than that of $L_*$ galaxies. By combining more massive star-forming galaxies from the literature with our MUSEQuBES sample, we observed that both $\left<N(\OVI)\right>$ and the average covering fraction peak at \logm $\approx9.5$, where the virial temperature is conducive for \OVI\ production via collisional ionization. We further argued that photoionization and/or non-equilibrium processes are necessary for the \OVI\ associated with low-mass dwarf galaxies (\logm~$<9$). Here, we explore the connection between galaxy properties and \OVI\ kinematics using a sample of 60 galaxy-absorber pairs with detected \OVI\ absorption.

This paper is organized as follows: In Section~\ref{sec:data} we describe the galaxy sample used in this paper, and the details of the absorption line data and galaxy-absorber pairs. The results are presented in Section~\ref{sec:results} followed by a discussion in Section~\ref{sec:discussion}. Our key findings are summarized in Section~\ref{sec:summary}. Throughout the paper, we adopt a $\Lambda$-CDM cosmology with $\Omega_{\rm m} = 0.3$, $\Omega_{\Lambda} = 0.7$, and a Hubble constant of $H_0 = 70~ \rm km~ s^{-1}~ Mpc^{-1}$. All distances are in physical units unless specified otherwise.

\color{black}

%%%%%%%%%%%%%%%%%%
\begin{figure*}
    \centering
    
    \includegraphics[width=0.5\linewidth]{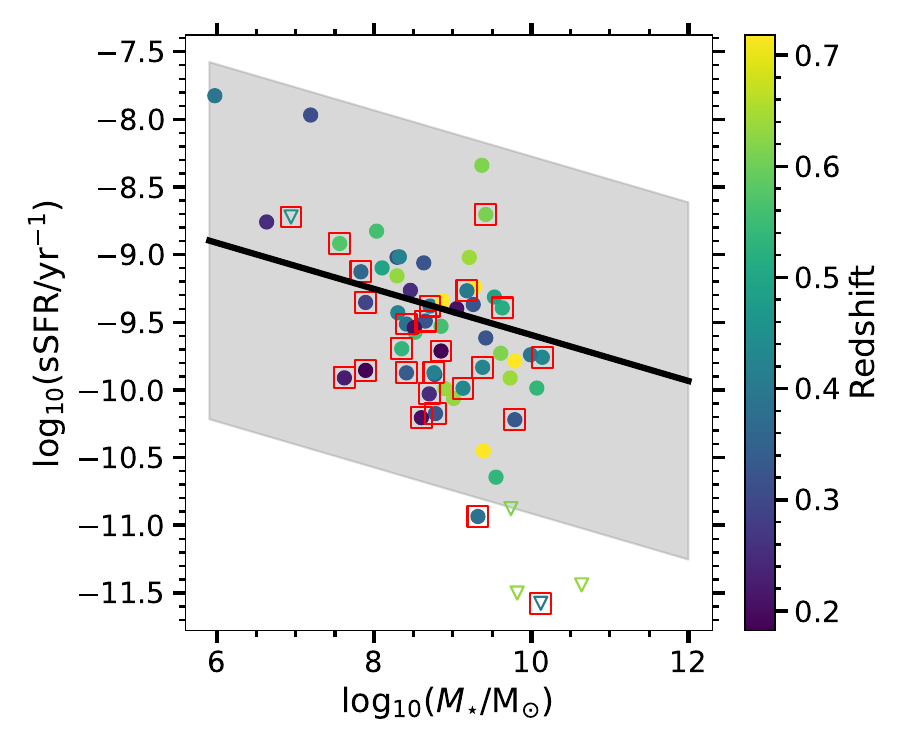}%
    \includegraphics[width=0.5\linewidth]{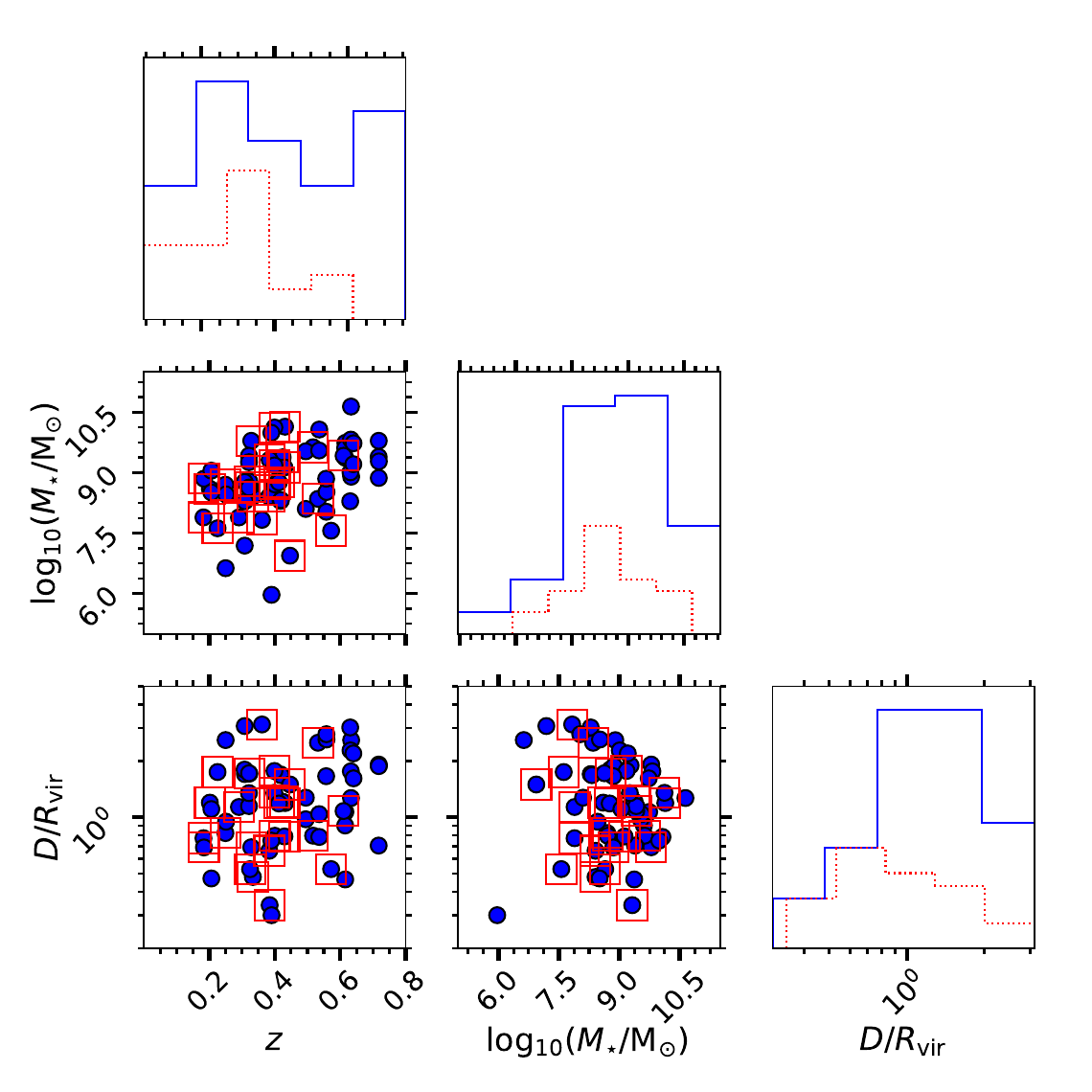}
    \caption{{\tt Left:} The sSFR of the 60 galaxies with detected \OVI\ absorption plotted against stellar mass. The filled circles and open triangles represent detections of and 3$\sigma$ upper limits on the sSFR, respectively. The points are color-coded by the redshift of the galaxies.  The SFMS relation at $z=0.5$ and its 3$\sigma$ scatter from \citet[]{Boogard_18} are shown with the black solid line and the gray shaded region, respectively. The square envelopes indicate `isolated' galaxies.
    {\tt Right:} The $D/R_{\rm vir}$, \logm, and $z$ distribution of galaxies with detected \OVI\ absorption plotted against each other, with histograms plotted along the diagonal panels. The full sample of galaxies is shown with filled blue circles and blue solid lines while the subsample of isolated galaxies is shown with red envelopes and red dotted lines.} 
    \label{fig:gal-prop}
\end{figure*}
%%%%%%%%%%%%%%%%%%

\section{Data} 
\label{sec:data}

We investigate the connection(s) between galaxy properties and the kinematics of circumgalactic \OVI-bearing gas. The absorber catalog and galaxy properties for a sample of 247 MUSEQuBES galaxies are taken from Paper I. We refer to \citet[]{Dutta_2024} for an overview of the survey design and Paper I for the results on the galaxy-absorber connection. Here, we briefly discuss the galaxy and absorber properties

\subsection{Galaxy survey}

The galaxies in the MUSEQuBES survey at low-$z$ are obtained from a blind spectroscopic redshift survey in fields centered on 16 quasars using the MUSE instrument on VLT/UT4. The 247 continuum-detected foreground galaxies have \OVI\ coverage in high quality $HST$/COS spectra of the quasars, free from contamination by geocoronal \lya\ or \OI\ emission. The redshifts of these galaxies are determined from emission (or absorption) features in the 1D spectra using MARZ \citep[]{Hinton_2016}, and further refined with a modified version of the code PLATEFIT \citep[]{Tremonti_2004}, which simultaneously fits Gaussian profiles to the available absorption and/or emission lines. 
Although the redshift returned by MARZ (resulting from galaxy template match) can be uncertain by $\approx100$ \kms, the refined redshift estimates of PLATEFIT are more robust (uncertainty in $z$ obtained from the fitting errors of the Gaussian fits in PLATEFIT is $\approx10$ \kms). 
Comparing the redshifts of galaxies from the MUSE observation of Hubble Ultra Deep Field in nine  $1'\times1'$  shallow {\tt MOSAIC} fields (exposure time $t_{\rm exp}\approx10$h which is somewhat higher than MUSEQuBES) with one $1'\times1'$ {\tt Ultra-Deep Field udf-10} (with $t_{\rm exp}\approx30$h), \citet[]{Inami_2017} showed that the uncertainty of spectroscopic redshfits observed with MUSE is $\approx40$ \kms. We can consider 40 \kms\ as the maximum redshift uncertainty of our galaxy sample for all practical purposes.

The star-formation rate (SFR) is measured from the H$\alpha$ or [\OII] fluxes  \citep[determined by PLATEFIT;][]{Tremonti_2004} using the relations in \citet[][for H$\alpha$]{Kennicutt_1998} or \citet[] [for \OII]{Kewley_2004} corrected to the \citet[]{Chabrier_2003} stellar initial mass function and corrected for dust (with H$\alpha$ and H$\beta$ line ratios). The stellar masses ($M_{\star}$) of these galaxies are obtained by the spectral energy distribution (SED) fitting \citep[using FAST][]{Kriek_2009} of the fluxes in 11 synthetic narrowband filters created from the full available wavelength range. Imposing a 3D Friends-of-Friend (FoF) algorithm with a linking velocity of $\pm500$ \kms\ and a linking separation of 500 kpc transverse distance enabled us to separate the galaxy sample into `isolated' and `group' subsamples. We use the term `group' to describe an association of two or more galaxies, without specifying the group halo mass, and we note that these structures may not necessarily be virialized.

\subsection{Absorber catalog}

The \OVI\ absorber identification in the 16 MUSEQuBES quasar fields is described in detail in Paper I. Briefly, a blind catalog of \OVI\ absorbers in the 16 quasar spectra was created through the visual inspection of the spectra using the doublet matching technique. The column densities of the detected absorbers are obtained with voigt-profile decomposition using {\sc vpfit} \citep[]{vpfit}. A total of 118 \OVI\ components are detected in the redshift range $0.13 < z_{\rm abs} < 0.71$. The best-fit redshift of the fitted components ($z_{\rm comp}$) and Doppler-$b$ parameters are used to characterize the \OVI\ bearing gas in the CGM of the galaxies. Using the component information, an absorption {\it system} is defined as the collection of individual absorption components located within $\pm300$~\kms\ from each other along the line-of-sight in a given sightline starting from the lowest redshift component. The 118 individual components can be grouped into
67 such systems\footnote{Using a different velocity window of $\pm200$ \kms\ and $\pm400$ \kms\ resulted in 67 and 65 absorption systems, respectively. Thus, our conclusions are not sensitive to the velocity window typically used to define $\Delta v_{90}$.}. The width of an absorption system is quantified by $\Delta v_{90}$, defined as the velocity extent within which 90\% of the total \OVI\ column density of the system is recovered.

The galaxy-absorber association is obtained by cross-matching our galaxy catalog with the blind \OVI\ absorber catalog with a linking velocity of $\pm$300 \kms\ centered on each galaxy redshift in a given MUSE field.
The column densities of all absorption components within the velocity windows are summed to obtain the total column density associated with a galaxy. The $z_{\rm comp}$ and $b$ of all the associated components are used for this study.
The redshift of an absorption system ($z_{\rm sys}$) in this context is defined as the column-density weighted mean redshift of the contributing components  {\footnote{ The grouping of components within $\pm$300 \kms\ centered on each galaxy redshift resulted in {\it systems} identical to the one described in the previous paragraph with the exception of two galaxies, each having one component of a {\it system} lying outside $\pm$300 \kms\ from the galaxy redshift.}}.

Out of the 247 galaxies used in this work, 60 exhibit associated \OVI\ absorption with a median of 2 components per system \footnote{Note that a single absorption system can be shared by multiple galaxies}.
In the left panel of Fig.~\ref{fig:gal-prop}, the specific star-formation rate (sSFR) of the galaxies is plotted against the stellar mass for the 60 galaxies with filled circles. The points are color-coded by the redshift of the galaxies. The black solid line represents the star-forming main sequence (SFMS) at $z=0.5$ \citep[]{Boogard_18}, with the grey shaded region indicating the 3$\sigma$ scatter around this relation.  
Consistent with the findings of \citet[]{Tumlinson_2011, Tchernyshyov_2022}, the galaxies associated with the detected \OVI\ absorptions are predominantly star-forming (within the 3$\sigma$ of SFMS). In Paper I, we show that the covering fraction of star-forming galaxies is significantly enhanced compared to passive galaxies inside the $R_{\rm vir}$. However, no significant difference in covering fraction is observed beyond $R_{\rm vir}$.

In this work, we present the kinematics of the \OVI\ bearing phase for the 60 galaxies with detected \OVI\ absorption, out of which 25 and 35 are classified as `isolated' and `group' galaxies, respectively. The right panel of Fig.~\ref{fig:gal-prop} shows the distribution of \logm, normalized impact parameter $D/R_{\rm vir}$ and $z$ with blue colors. The isolated galaxy properties are indicated with red envelopes in both panels. The red-dotted histograms represent the distribution of isolated galaxy properties in the right panel.

\section{Results} 
\label{sec:results}

%\subsection{\OVI\ kinematics}

%\blue{
%We investigate the connection between the host galaxy properties and the kinematics of \OVI\ absorption with the aid of MUSEQuBES galaxy-absorber pairs. }

\begin{figure}
    \centering
    \includegraphics[width=1\linewidth]{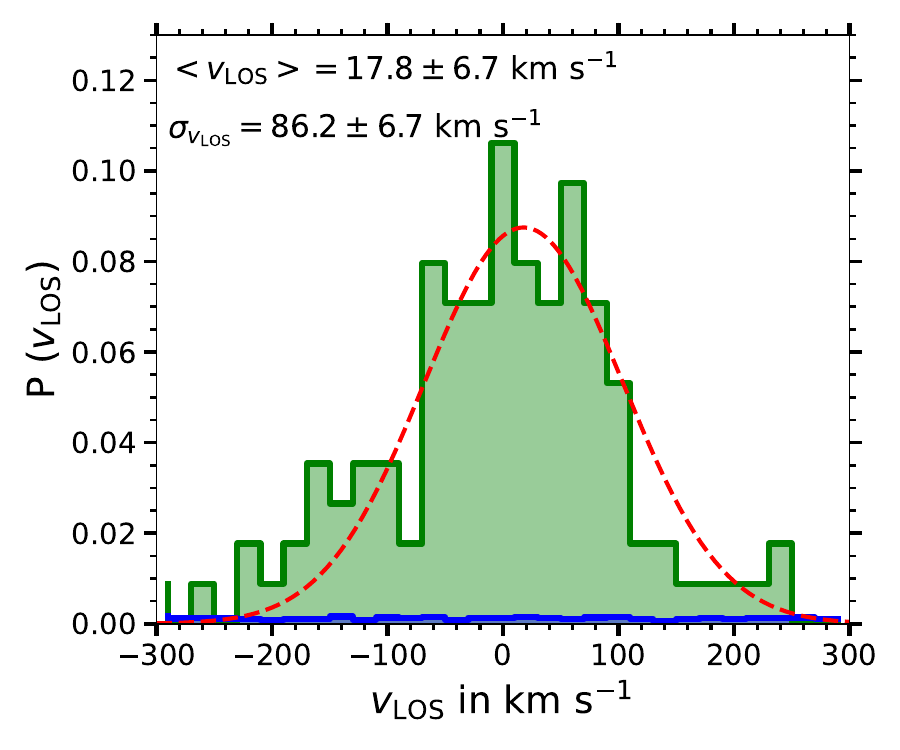}
    \caption{ LOS velocity distribution of \OVI\ components with respect to the host galaxy redshift, normalized by the number of galaxy-component pairs in the MUSEQuBES sample. The filled green histogram represents the \OVI\ components around galaxies. The solid blue histogram represents the average absorber density near randomly chosen redshifts drawn from our galaxy redshift distribution. The red dashed line represents the best-fit Gaussian to the observed LOS velocity distribution, with the best-fit mean ($\left<v_{\rm LOS}\right>$) and standard deviation ($\sigma_{v_{\rm LOS}}$) indicated inside the panel.
    }  
    % Probability density of LOS velocity separation of \OVI\ components with respect to the host galaxy redshift shown with green histograms. The blue-hatched histograms represent the same but for 20000 random redshifts. The red dashed line represents the best-fit Gaussian profile to the observed $v_{\rm LOS}$ distribution. The mean and standard deviation of the best-fit profile are indicated inside the panel. }
    \label{fig:dv_all}
\end{figure}

%%%%%%%%%%%%
\begin{figure*}
    \centering
    \includegraphics[width=0.5\linewidth]{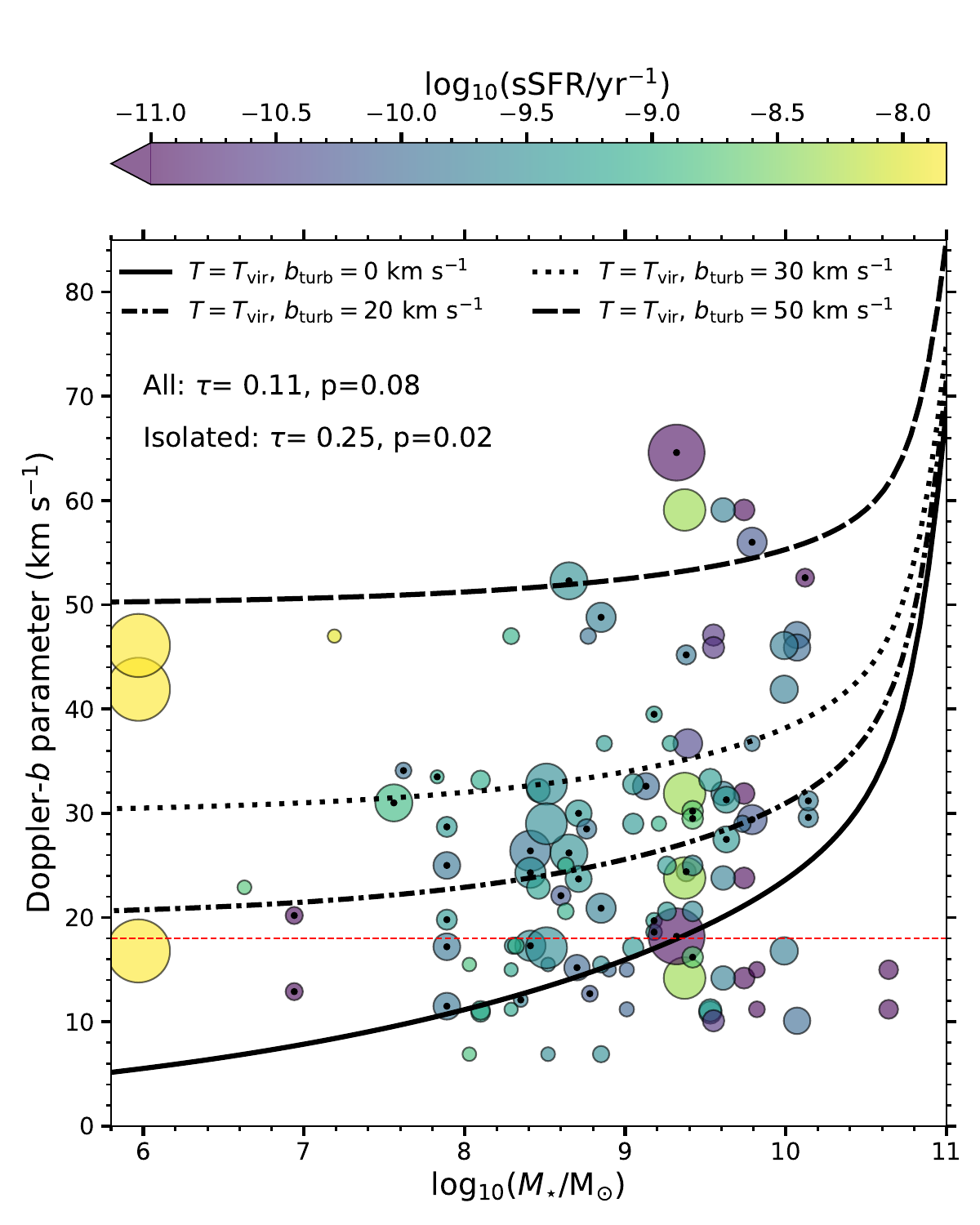}%
    \includegraphics[width=0.5\linewidth]{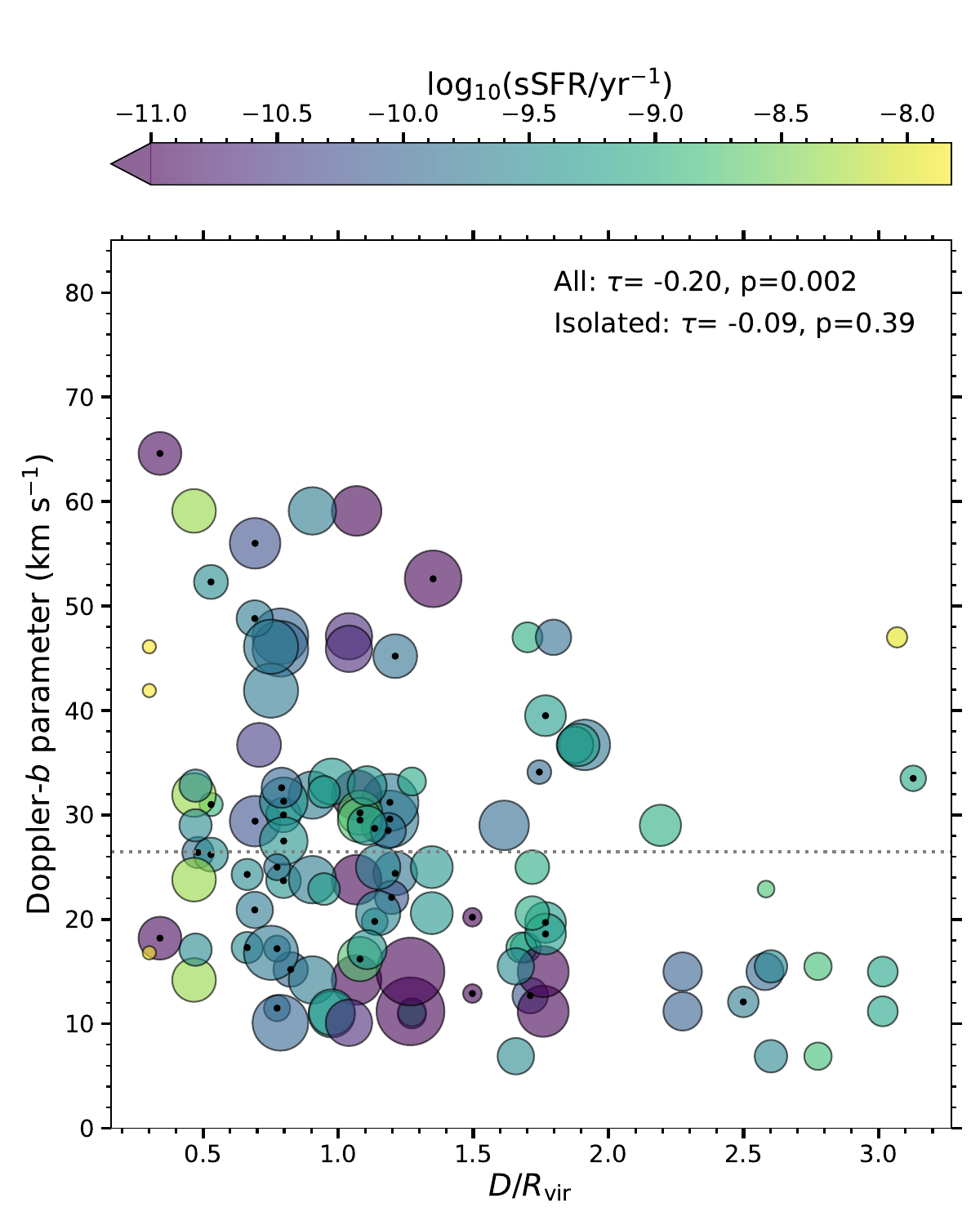}
    \caption{{\tt Left:} $b$-parameter of \OVI\ components plotted against the stellar mass of the host galaxy. The circles are color-coded by the sSFR. The galaxies with upper limits on the sSFR are assigned the flag value of $10^{-11}~{\rm yr}^{-1}$, where the color scale saturates. The area of the circles is proportional to the $(D/R_{\rm vir})^{-2}$ of the host galaxy. The absorption components hosted by isolated galaxies are marked with black dots in the center. The solid, dash-dotted, dotted, and dashed lines represent the expected $b$-parameter for the virial temperature of the host halo along with the turbulent velocities of 0, 20, 30, and 50 \kms, respectively. The horizontal red dashed line indicates the thermal width corresponding to the temperature at which the \OVI\ ion-fraction peaks in collisional ionization. {\tt Right:} $b$-parameter plotted against $D/R_{\rm vir}$ of the host galaxy. The color-coding is the same as on the left. The area of the circles is proportional to the stellar mass of the host galaxy. The horizontal gray dotted line indicates the median $b$ value of \OVI\ components with ${\rm log}_{10}(N(\OVI)/{\rm cm}^{-2})<14$ from \citet[]{Danforth_2016}. }
    \label{fig:b-m_all}
\end{figure*}

%%%%%%%%%%%%

\subsection{LOS velocity of absorbing gas w.r.t. the galaxy}  

The distribution of LOS velocity offset ($v_{\rm LOS}$) between the velocity centroid of \OVI\ components with respect to the host galaxy redshift is shown in Fig.~\ref{fig:dv_all} with the green histogram constructed with bins of 20 \kms. Here, $v_{\rm LOS}$ is given by

\begin{equation}
    v_{\rm LOS}=(3\times10^{5}~{\rm km~ s^{-1})} 
     \times \frac{ z_{\rm comp}-z_{\rm gal} } {1+z_{\rm comp}} .
\end{equation}

We define ${\rm P}(v_{\rm LOS})$ as the ratio of the number of \OVI\ components in a given $v_{\rm LOS}$ bin to the total number of galaxy-absorber pairs in the sample. 
The red dashed line represents the best-fit Gaussian profile to the observed $v_{\rm LOS}$ distribution. The mean ($\left<v_{\rm LOS}\right>$) and standard deviation ($\sigma_{v_{\rm LOS}}$) obtained from the best-fit Gaussian profile are  $\approx18\pm7$ and $\approx86\pm7$~\kms, respectively.
The blue histograms represent ${\rm P}(v_{\rm LOS})$ for 20000 randomly chosen redshifts drawn from our galaxy redshift distribution.
Strongly enhanced velocity clustering of absorbers relative to the random region is observed within $\approx\pm200$ \kms\ from the host galaxy redshift.

Next, we repeat this exercise for \OVI\ {\em systems} in which the absorption velocity centroid is determined from the column density-weighted mean redshift of the associated components. Such an analysis produces consistent results ($\left<v_{\rm LOS}^{\rm sys}\right> \approx16\pm12~$\kms, $\sigma_{v_{\rm LOS}^{\rm sys}}\approx 77\pm12~$\kms).

It is worthwhile to mention that the 60 galaxies with detected \OVI\ absorption in our MUSEQuBES sample have a median \logm$=8.9$, corresponding to a virial velocity of $\approx80$ \kms\ at $z=0.4$. The observed $\sigma_{v_{\rm LOS}}$ is higher than the line-of-sight component of the virial velocity ($\sigma^{\rm vir}_{\rm LOS} = 80/\sqrt{3} \approx 46$ \kms). Even considering the redshift uncertainties ($\Delta z$) of $\approx40$ \kms, the total extent due to the virial velocity and redshift uncertainties $\left(\Delta z^2 + (\sigma^{\rm vir}_{\rm LOS})^2 \approx 60~{\rm km~s^{-1}} \right)$ is marginally lower than the observed $\sigma_{v_{\rm LOS}}$.

Dividing our galaxy-absorber samples into two $D/R_{\rm vir}$ bins, we observe a significantly higher $\sigma_{v_{\rm LOS}}\approx100$ \kms\ ($\sigma_{v_{\rm LOS}^{\rm sys}}\approx87$ \kms) for galaxies at $D>R_{\rm vir}$, compared to $\sigma_{v_{\rm LOS}}\approx76$ \kms\ ($\sigma_{v_{\rm LOS}^{\rm sys}}\approx70$ \kms) for galaxies at $D\leq R_{\rm vir}$. The increase in $\sigma_{v_{\rm LOS}}$ at larger $D/R_{\rm vir}$ can be a consequence of an increasing number of background/foreground or correlated absorbers outside the halo. 
Restricting to the galaxies with $D\leq R_{\rm vir}$, we do not observe any significant difference of $\sigma_{v_{\rm LOS}}$ between high-- (\logm~$\geq9$) and low--mass (\logm~$<9$) galaxies. We note that the observed $\sigma_{v_{\rm LOS}^{\rm sys}}\approx70$ \kms\ within $R_{\rm vir}$ for the \OVI--bearing gas is consistent with the dispersion of \MgII--bearing gas around \logm~$\approx9.6$ galaxies reported by \citet[]{Huang_2021}, despite the fact that \OVI\ and \MgII\ are thought to trace two different phases of the CGM.

\subsection{Doppler $b$-parameter}
\label{sec:dop-b}

The internal kinematics of the \OVI\ absorption can be quantified by the Doppler $b$-parameter of individual Voigt components. In the left panel of Fig.~\ref{fig:b-m_all}, we plot the $b$-parameter of individual components against the stellar mass of host galaxies. The circles are color-coded by the sSFR of the host galaxy. The galaxies with upper limits on the sSFR are assigned the flag value\footnote{The actual upper limits can be higher than the flag value} of $10^{-11}~{\rm yr}^{-1}$, where the color scale saturates. Additionally, the area of the circles is proportional to the $(D/R_{\rm vir})^{-2}$ of the host galaxy. The points with small black dots in the center indicate the components associated with `isolated' galaxies in our sample. The Kendall-$\tau$ correlation coefficient indicates a weak correlation ($\tau=0.11, p = 0.08$) between the $b$-parameter and the stellar mass ($M_{\star}$) for the full sample, with a stronger correlation observed when considering measurements for `isolated' galaxies ($\tau=0.25, p = 0.02$) only. Notably, the broad components with $b>40$~\kms\ are primarily associated with relatively massive galaxies. However, massive galaxies also exhibit components with low $b$ values.

The $b$-parameter includes contributions from both thermal and non-thermal or turbulent motion. A robust upper limit on the temperature can be inferred from $b$, as 
\begin{equation}
    b^2=2kT/m_{\rm O}+ b_{\rm turb}^2~, 
\end{equation}
where $b_{\rm turb}$ represents the contribution from non-thermal motions and $k$ and $m_{\rm O}$ are the Boltzmann constant and mass of the oxygen atom, respectively. Assuming $b_{\rm turb} = 0$, conservative upper limits on the temperature can be obtained from the $b$ parameter as $T_{\rm max}= 1.5 \times 10^{6} \left(\frac{b}{40~{\rm km~s^{-1}}}\right)^2~ \rm K$. For galaxies with \logm~$<10$, the virial temperature is $<10^6$~K. The inferred $T_{\rm max}$ for $b=40$~\kms\ is $\approx10^6$~K, implying that the maximum allowed gas temperature is consistent with the virial temperature of the galaxies' host haloes.

In the left panel of Fig.~\ref{fig:b-m_all}, we show the expected \OVI\ $b$-parameter for gas at the virial temperature of the halo with $b_{\rm turb}$ values of 0, 20, 30, and 50 \kms\ in solid, dash-dotted, dotted, and dashed lines, respectively. The virial temperature of a galaxy is determined from the halo mass, obtained using the abundance-matching relation of \citet[]{Moster_2013}. A significant fraction of the \OVI\ components exhibits $b$ values that are consistent with $b_{\rm turb}\geq20$~\kms, assuming the maximum attainable gas temperature is equal to the virial temperature. We will discuss turbulence in the CGM further in Section~\ref{disc:turb}.

The red dashed line in Fig.~\ref{fig:b-m_all} indicates the $b$-parameter corresponding to the temperature at which the \OVI\ ion-fraction peaks ($b_{\rm peak}$) in collisional ionization equilibrium (i.e., $T_{\rm peak} = 10^{5.5}$~K). 19 out of the 113 components ($\approx 17$\%) exhibit $b+\Delta b < b_{\rm peak}$, where $\Delta b$ is the $1\sigma$ error in $b$ returned by {\sc vpfit}. This suggests that the absorbing gas for these components is cooler than the temperature required for CIE. Interestingly, 16 out of these 19 components are detected at $D \geq R_{\rm vir}$, as indicated by their small sizes. Moreover, out of the 20 components with $b+\Delta b$ below the solid black line, indicating the maximum thermal broadening for the virial temperature ($b(T_{\rm vir})$), 14 are associated with galaxies with $D \geq R_{\rm vir}$. Clearly, the majority of the \OVI\ components are broad enough to accommodate the temperatures required for collisional ionization. The small fraction for which the $b$ value is too low for \OVI\ to be collisionally ionized is predominately located at $D\geq R_{\rm vir}$. In the outer CGM, such absorption components may arise from photoionization by the extragalactic UV background radiation.

At this point, we reiterate that a single \OVI\ component may be associated with multiple group galaxies, resulting in counting the same component multiple times. Associating the galaxy with the smallest $D/R_{\rm vir}$ in a given group with the \OVI\ components results in 8 components with $b+\Delta b < b(T_{\rm vir})$.  4 out of the 8 components are hosted by galaxies with $D/R_{\rm vir}<1$, and all of them have \logm$\geq9.4$.

In the right panel of Fig.~\ref{fig:b-m_all}, we plot the $b$-parameter of individual components against the $D/R_{\rm vir}$ of the associated galaxies. The size of the circles is proportional to the stellar mass of the host galaxy. Other details are similar to the left panel of Fig.~\ref{fig:b-m_all}. A weak anti-correlation is observed between $D/R_{\rm vir}$ and $b$ with a Kendall-$\tau$ correlation coefficient of $\tau=-0.19$ ($p=0.002$) for the full sample. No such correlation is seen for the isolated subsample ($\tau=0.09, p= 0.39$). It is evident that the large $b$-values ($b>40$~\kms) are seen only within $\approx2R_{\rm vir}$.  The dotted horizontal line in the plot shows the median $b$-parameter ($\approx 27$~\kms) of the \OVI\ components in the ``galaxy-blind'' catalog of \citet[]{Danforth_2016} with  ${\rm log}_{10}(N(\OVI)/{\rm cm}^{-2})<14$. This can be considered the median \OVI\ $b$-parameter in random regions.
The median $b$-value of the components within $\approx 2R_{\rm vir}$ is 26~\kms\ whereas it is 15~\kms\ at $>2R_{\rm vir}$. It is interesting to note that the majority of the galaxies at $>2R_{\rm vir}$, which likely drives the weak anti-correlation seen for the full sample, are not isolated. A 2-sample KS test reveals a marginally different $b$-distribution for the isolated and group galaxies ($p=0.057$), albeit a comparable median $b$ of $\approx27$ and 24 \kms, respectively. Restricting to galaxies with $D<R_{\rm vir}$, however, we do not observe any statistical difference in the $b$ distributions of the isolated and group galaxies ($p=0.196$). Finally, there is no significant color gradient in the plot, indicating that there is no evidence for a correlation between the sSFR and the gas kinematics.

%%%%%%%%%%%%
\begin{figure}
    \centering
    \includegraphics[width=0.98\linewidth]{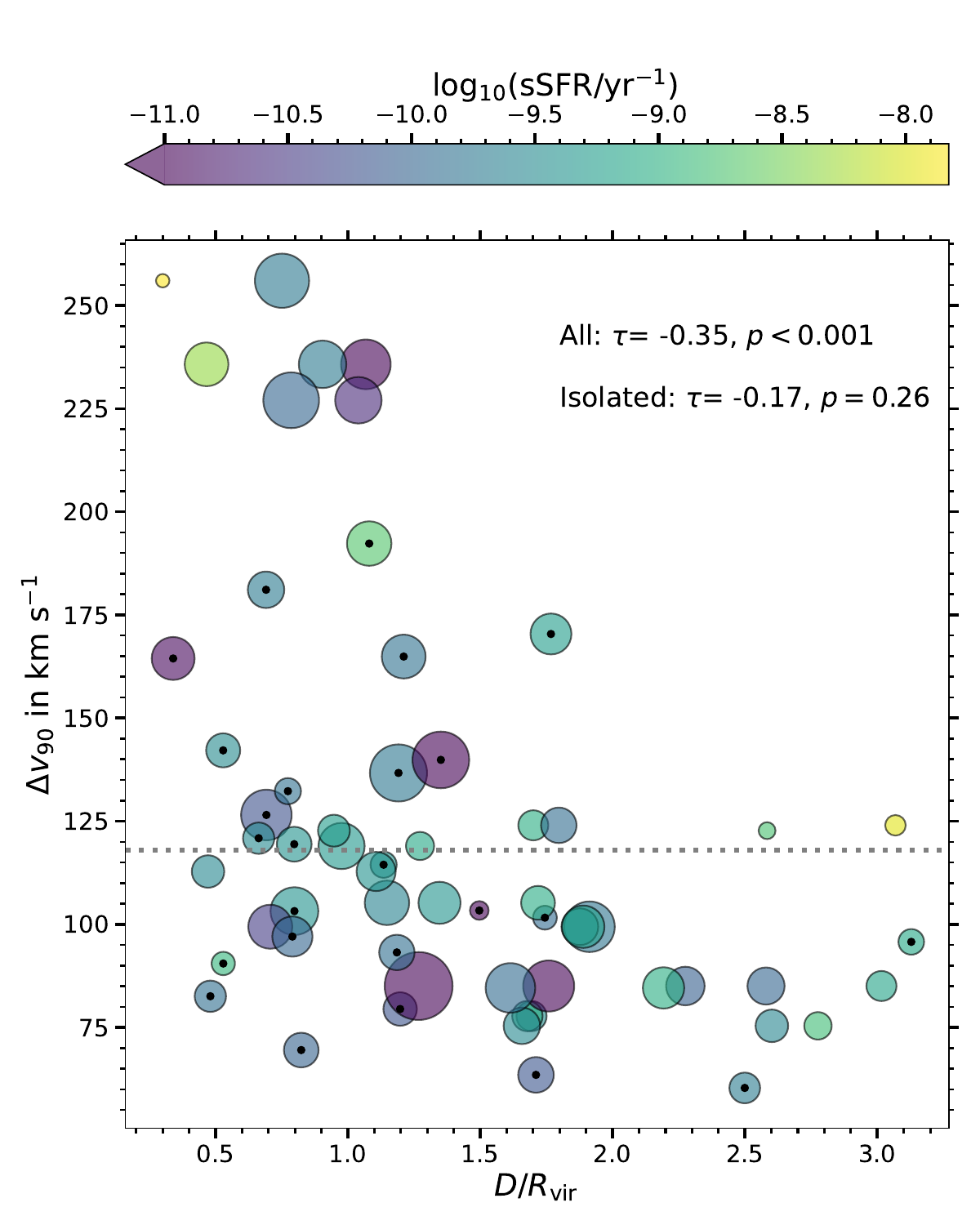}
    \caption{$\Delta v_{90}$ of \OVI\ absorption systems plotted against the $D/R_{\rm vir}$ of the associated galaxy. The points are color-coded by the sSFR of the galaxy. The galaxies with upper limits on sSFR are assigned the flag value of $10^{-11}~{\rm yr}^{-1}$, where the color scale saturates. The area of the points is proportional to the stellar mass of the host galaxy. The absorption systems hosted by isolated galaxies are marked with black dots in the center. The horizontal gray dotted line indicates the median $\Delta v_{90}$ constructed from the `galaxy-blind' \OVI\ catalog of \citet[]{Danforth_2016}, considering only systems without any component with ${\rm log}_{10}(N(\OVI)/{\rm cm}^{-2})\geq 14$.}
    \label{fig:v90_dn}
\end{figure}
%%%%%%%%%%%%

\subsection{$\Delta v_{90}$ and TPCF}
\label{sec:tpcf&v90}

%%%%%%%%%%%%%%
\begin{figure*}
    \centering
    \includegraphics[width=1\linewidth]{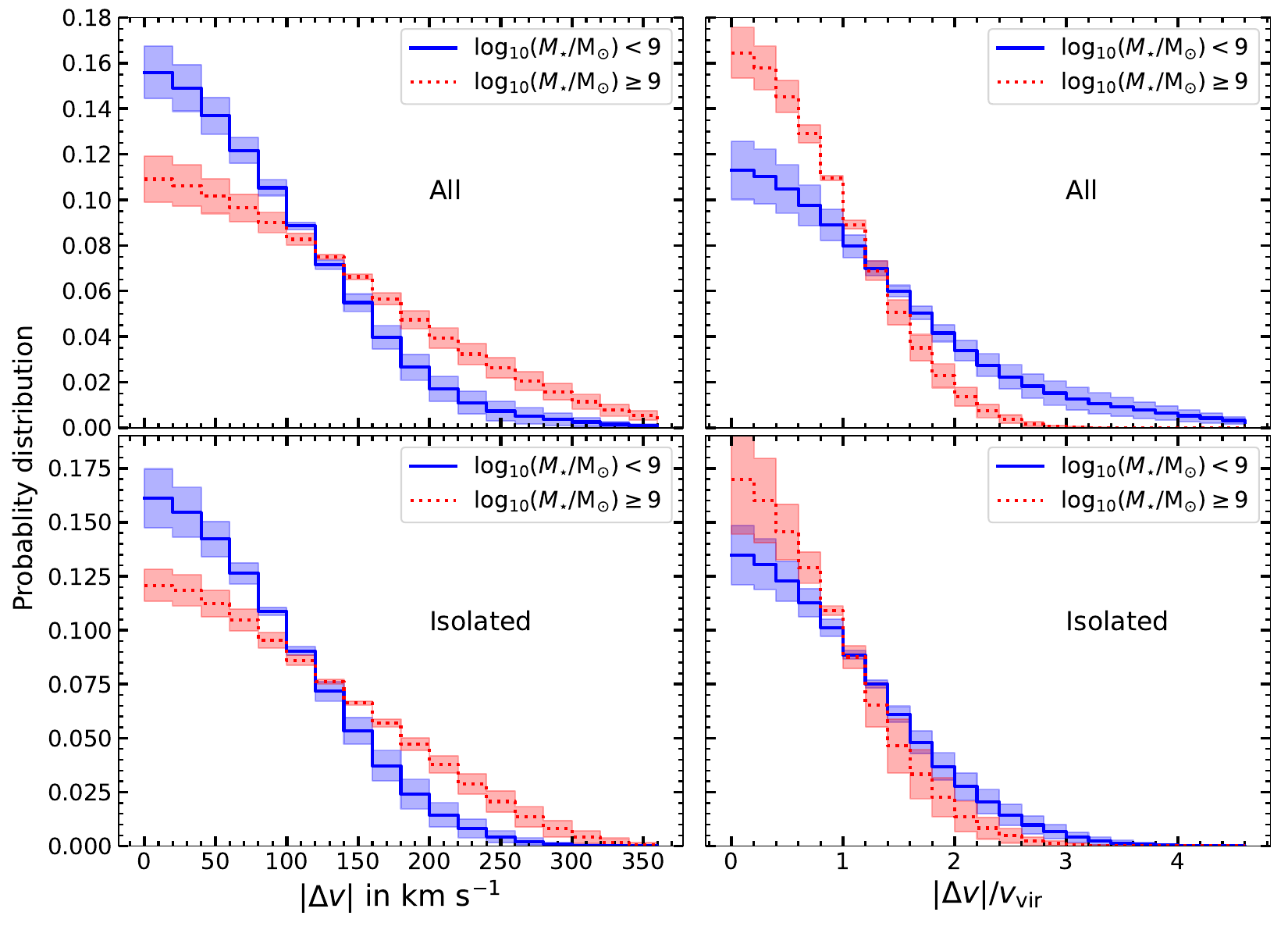} 
    \caption{Comparison of the pixel velocity two-point correlation function (TPCF) for low-mass (\logm$<9$), and high-mass (\logm$\geq9$) bins, shown with solid blue, and dotted red histograms, plotted against $|\Delta v|$ in the left panels and $|\Delta v|/v_{\rm vir}$ in the right panels. The top and bottom rows show the TPCF for the full sample and the isolated subsample, respectively. }
    \label{fig:tpcf_m_all}
\end{figure*}
%%%%%%%%%%%%%

While the $b$-parameter provides insights into the kinematics of individual components, the inter-cloud kinematics is better captured by the full kinematic extent of the absorption systems. The full kinematic extent of an \OVI\ absorber can be quantified using $\Delta v_{90}$ and the pixel-velocity two-point correlation function (TPCF).

We define $\Delta v_{90}$ as the velocity range that encompasses 90\% of the total column density of the system. In Fig.~\ref{fig:v90_dn}, we plot $\Delta v_{90}$ against $D/R_{\rm vir}$ of the associated galaxy. The circles are color-coded by the sSFR of the galaxy. The galaxies with upper limits on the sSFR are assigned the flag value of $10^{-11}~{\rm yr}^{-1}$, where the color scale saturates. Additionally, the area of the circles is proportional to the stellar mass of the host galaxy. The absorption systems associated with isolated galaxies are marked with black dots in the center. The Kendall-$\tau$ correlation coefficient reveals a moderate anti-correlation ($\tau=-0.35, p < 0.001$) between the $\Delta v_{90}$ and $D/R_{\rm vir}$ for the full sample of 60 galaxies, which vanishes when only the 25 isolated galaxies are considered. The anti-correlation is primarily driven by the high $\Delta v_{90}$ systems associated with galaxies at $D/R_{\rm vir}<1$. The broad \OVI\ systems ($\Delta v_{90}>200$~\kms) are primarily associated with galaxies in groups. 
% Besides, the overall decrease in the symbol size from top to bottom of the plot implies narrower kinematic extents for the low-mass galaxies.
A Kendall-$\tau$ correlation yields a tentative, weak correlation between \logm\ and $\Delta v_{90}$ with $\tau=-0.20, ~p=0.2$ ($\tau=-0.17,~p=0.09$) for galaxies with $D/R_{\rm vir}<1$ ($D/R_{\rm vir}<2$).

The gray horizontal dotted line in Fig.~\ref{fig:v90_dn} indicates the median $\Delta v_{90}$ of the \OVI\ systems in the catalog of \citet[]{Danforth_2016}. To obtain this information, we first used the Voigt profile fit parameters of all the \OVI\ components in their catalog and defined systems using a linking velocity of 300~\kms. We excluded the systems in which one of the components had ${\rm log}_{10}(N(\OVI)/{\rm cm}^{-2})\geq 14$ as these strong absorbers are likely to be associated with the CGM of galaxies\footnote{Including these systems do not result in appreciable change of the median $\Delta v_{90}$ of random regions.}. For each selected system, we generated a model absorption profile using the wavelength-dependent LSF of COS. Such model profiles are then used to determine the $\Delta v_{90}$.

The median $\Delta v_{90}$ of $\approx120$ \kms\ from \citet[]{Danforth_2016} is comparable to the median $\Delta v_{90}\approx123$~\kms\ of the \OVI\ systems associated with galaxies at $D<R_{\rm vir}$. However, the median $\Delta v_{90}\approx227$~\kms\ for the group galaxies with $D<R_{\rm vir}$ is significantly higher than for the random regions. A 2-sample KS test between the $\Delta v_{90}$ distribution of group and isolated galaxies with $D<R_{\rm vir}$ results in a $p$-value of $0.06$. This marginal difference is not present when galaxies at all impact parameters are considered ($p=0.3$). The galaxies with $D\geq R_{\rm vir}$ have marginally lower median $\Delta v_{90}\approx100$~\kms. Similar to the analysis of the $b$-parameters, we find that the majority of the galaxies with narrow $\Delta v_{90}$ at $D> R_{\rm vir}$ are not isolated.

Further binning both the isolated and group subsamples based on stellar mass, we find a marginally higher $\Delta v_{90}\approx140$~\kms\ for the massive (\logm$\geq9$) isolated galaxies compared to the low-mass (\logm$<9$) isolated galaxies $\Delta v_{90}\approx100$ \kms. A significant difference in the median $\Delta v_{90}$ between the low- and high-mass group galaxies ($\Delta v_{90}\approx 123~ {\rm and}~230$ \kms, respectively) is observed only when the impact parameter is $<R_{\rm vir}$. $\Delta v_{90}$ is low for group galaxies at $D\geq R_{\rm vir}$ irrespective of the stellar mass.

Next, we used the pixel-velocity TPCF as an alternate method to characterize the velocity dispersion of \OVI\ absorption systems \citep[see e.g.,][]{Nielsen_2015}. Briefly, we first compile all the pixel velocities (with respect to the galaxy redshift) within the full extent of the absorption systems for a given subsample of galaxies. The absolute velocity differences of all pixel pairs are calculated from this compilation before binning them into 20~\kms\ velocity windows. The count in each bin is then normalized by the total number of pairs in the subsample to obtain the probability distribution. We divide our MUSEQuBES galaxy sample with detected \OVI\ absorption into 2 bins with \logm~$<9$ and \logm~$\geq 9$. There are 33 and 27 galaxies in each respective bin. The \OVI\ TPCFs for low- and high-mass bins are shown in the top left panel of Fig.~\ref{fig:tpcf_m_all} with blue and red histograms. The shaded regions are the bootstrap errors obtained from 100 bootstrap realizations, with absolute velocity differences calculated from resampled pixel velocities. The \OVI\ TPCF for the high-mass bin is significantly wider compared to its low-mass counterpart.

The larger velocity dispersion of \OVI\ absorption in high-mass halos can arise from their higher circular velocities. In order to account for this difference, we produced a new \OVI\ TPCF after scaling the absolute velocity differences of pixel pairs of an absorption system by the virial velocity ($\sqrt{GM_{\rm vir}/R_{\rm vir}}$) of the associated galaxy. The resulting scaled TPCFs for the low- and high-mass galaxies in the MUSEQuBES sample are shown in the top right panel of Fig.~\ref{fig:tpcf_m_all} with blue and red histograms, respectively. Interestingly, the scaled \OVI\ TPCF reveals a reverse trend, with the low-mass galaxies showing significant probabilities for velocities larger than the virial velocity. The profile for the high-mass galaxies, on the other hand, shows a sharp decline beyond the virial velocity. The excess probability at $|\Delta v|/v_{\rm vir}>1$ for low-mass galaxies may be due to material efficiently escaping the halos with shallower gravitational potential wells.

However, another possible explanation for this excess probability is the presence of neighboring galaxies that share the same absorption system. To disentangle any potential role of the environment in the mass trend discussed above, we repeated the analysis using a subsample of only isolated galaxies. The \OVI\ TPCF and normalized \OVI\ TPCF for the isolated galaxies are shown in the bottom-left and bottom-right panels of Fig.~\ref{fig:tpcf_m_all}.

Consistent with the full sample, we observe a broader kinematic extent for the absorbers hosted by more massive galaxies (bottom-left panel). However, this difference largely vanishes for the normalized \OVI\ TPCF (bottom-right panel). There is only a marginal $2.3\sigma$ difference observed for the two mass bins when the environment is controlled, as opposed to the $\approx20 \sigma$ difference seen for the full sample (top-right panel). The decrease in significance can be attributed to the smaller sample size. Alternatively, this may indicate that the observed difference is primarily driven by the environment. The role of the environment in the kinematic extent of \OVI\ absorption is further discussed in Sec.~\ref{disc:kinematics_width}.

\section{Discussion}
\label{sec:discussion}

We have studied the connection between galaxy properties and the kinematics of \OVI\ absorption in the CGM of low-redshift ($0.1-0.7$) galaxies.

In this section, we discuss the main findings from Section~\ref{sec:results}.

%%%%%%%%%%%% 
\begin{figure*}
    \centering
    \includegraphics[width=0.75\linewidth]{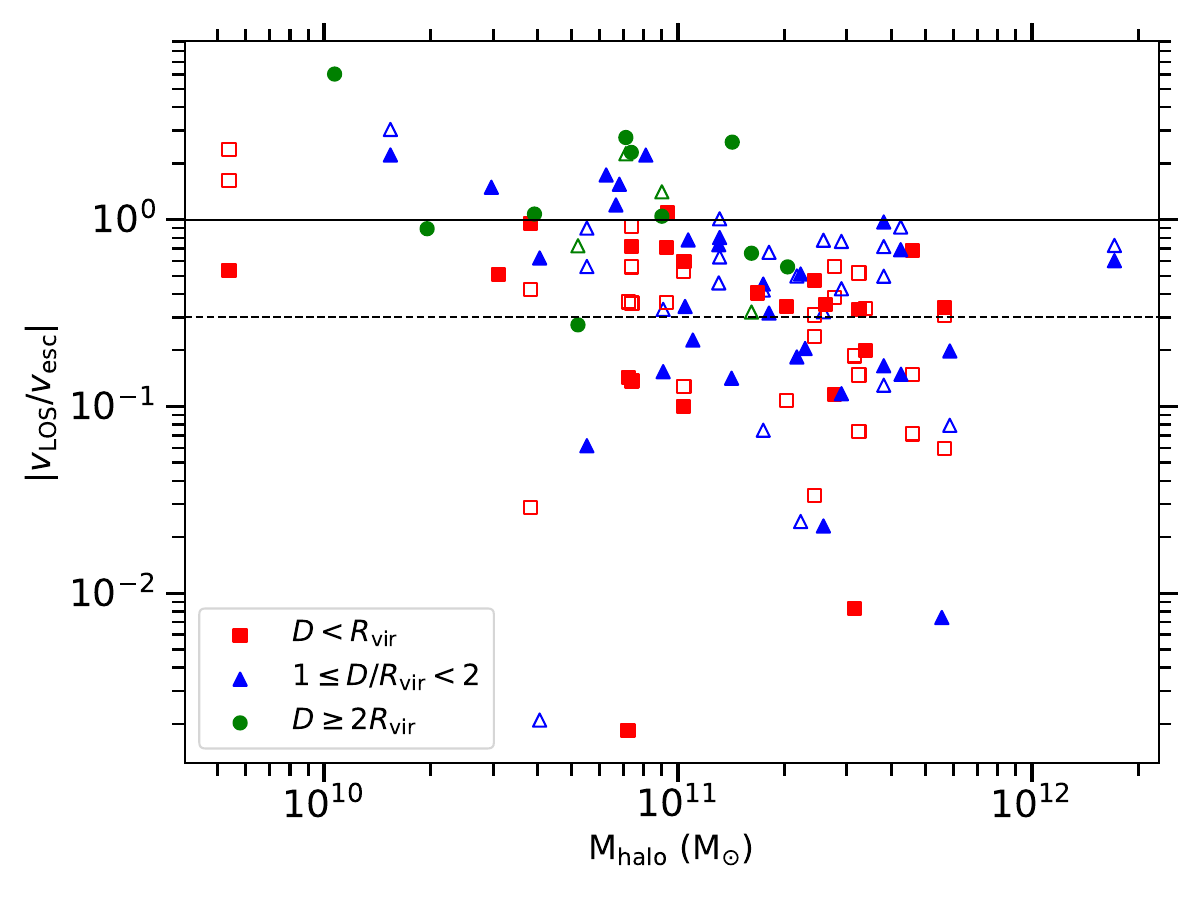}   
    \caption{The velocity of \OVI\ components ($v_{\rm LOS}$) scaled by local escape velocity $\left[v_{\rm esc}(D)\right]$ is plotted against the halo mass of the associated galaxy for the MUSEQuBES sample with red squares (for galaxies with $D<R_{\rm vir}$), blue triangles ($1\leq D/ R_{\rm vir}<2$) and green circles ($D\geq 2R_{\rm vir}$). The solid and hollow markers represent the primary (strongest) and secondary components of an absorption system. The horizontal lines indicate $|v_{\rm LOS}/v_{\rm esc}(D)|=1$ (solid) and $|v_{\rm LOS}/v_{\rm esc}(D)|=1/\sqrt{3}$ (dashed) (the latter accounts for projection effect). }
    \label{fig:dv_mhalo_esc}
\end{figure*}
%%%%%%%%%%%%

%%%%%%%%%%%%%%
\begin{table*}
\centering
\caption{Summary of \OVI\ bound fraction measurements}   
\label{tab:OVI_bound_frac}
\begin{tabular*}{\textwidth}{l@{\extracolsep{\fill}}cccc} 
\hline

 ${\rm log}_{10}(M_{\rm halo}/{\rm M_{\odot}})$   & \multicolumn{4}{c}{Upper limit on the bound fraction at different $D/R_{\rm vir}$}  \\ \cline{2-5} 
  & All & $D<R_{\rm vir}$ & $1\leq D/ R_{\rm vir}<2$ & $D\geq 2R_{\rm vir}$  \\ \hline 
  \hline
 All  & 0.83 (0.62) & 0.94 (0.81) & 0.85 (0.56)  &  0.43 (0.21)  \\ 
 $\leq 11.5$ & 0.78 (0.59) & 0.91 (0.77) & 0.79 (0.56) & 0.43 (0.21)  \\ 
 $> 11.5$ & 1.0 (0.72) & 1.0 (0.92) & 1.0 (0.54) & -  \\ 
 \hline 
\end{tabular*}
 \justify  
 \item Notes--\begin{itemize}
     \item There are no galaxies with \logm$>11.5$ at $D\geq2R_{\rm vir}$.
     \item The bound fraction w.r.t the projected escape velocity $(v_{\rm esc}(D)/\sqrt{3})$ is indicated inside parentheses.  
 \end{itemize}  %% Bound fraction is defined as the ratio $n_1/n_1+n_2$, where $n_1$ and $n_2$ are the numbers of components with $|v_{\rm LOS}/v_{\rm esc}|<1$ and  $|v_{\rm LOS}/v_{\rm esc}|\geq 1$, respectively. 
\end{table*}
%%%%%%%%%%%%

\subsection{Are the {\rm O~{\sc vi}} components bound to the dwarf galaxies?} 

The velocity distribution of \OVI\ absorption components around MUSEQuBES galaxies shows significant clustering within $\pm86$~\kms\ (Fig.~\ref{fig:dv_all}). The velocity scale of clustering is larger\footnote{Even when the redshift uncertainties of $\approx40$ \kms\ are considered.} than the projected virial velocity of $\approx46$~\kms\ for galaxies with the median \logm~$=8.9$ in our MUSEQuBES sample for which \OVI\ absorption is detected.
Here we investigate how the line of sight velocities compare with respect to the local escape velocities of the halos.

In Fig.~\ref{fig:dv_mhalo_esc}, we show the absolute LOS velocity of individual \OVI\ components w.r.t. the host galaxy redshift, normalized by the local escape velocity\footnote{We assumed a NFW density profile for the host halo mass, with local escape velocity at a given distance $D$ given by $v_{\rm esc}(D)=\sqrt{(2GM_{\rm halo}/D) {\rm ln}[1+c(D/R_{\rm vir})]/[{\rm ln}(1+c)-c/(1+c)]} $. The halo-mass and redshift-dependent concentration parameter $c$ is obtained using the
% {\sc Colossus} package \citep[]{Diemer_2018}.
{\sc Commah} package \citep[]{Correa_2015}} ($|v_{\rm LOS}/v_{\rm esc}(D)|$), plotted against the halo mass with red squares (for galaxies with $D<R_{\rm vir}$), blue triangles (for galaxies with $1\leq D/R_{\rm vir}<2$), and green circles (for galaxies with $D\geq 2R_{\rm vir}$) for the galaxies in our sample. The solid and hollow symbols represent the primary (strongest) and secondary (weaker) components of an absorption system. The horizontal solid line represents $|v_{\rm LOS}/v_{\rm esc}(D)|= 1$. The horizontal dashed line represents $|v_{\rm LOS}/v_{\rm esc}|= 1/\sqrt{3}$ which accounts for the projection effects on the LOS velocity.

We determine the ``bound fraction'' defined as $n_1/(n_1+n_2)$, where $n_1$ and $n_2$ are the number of components with $|v_{\rm LOS}/v_{\rm esc}(D)| \leq 1$ and $>1$, respectively. However, we note that this number relies on quantities with large uncertainties (such as $M_{\rm halo}$). Additionally, the impact parameter $D$ represents the {\it minimum} possible 3D distance of the absorbers and $v_{\rm LOS}$ is a lower limit on the 3D velocity. Hence, this is an upper limit on the fraction of truly bound absorbers. The bound fractions for different halo mass and impact parameter bins are tabulated in Table \ref{tab:OVI_bound_frac}. The bound fraction is $<$83\% for the full MUSEQuBES sample, which reduces to $<$62\% when the $|v_{\rm LOS}/v_{\rm esc}(D)| \leq 1/\sqrt{3}$ is considered to be bound. This indicates that, overall, a considerable fraction of the \OVI\ components in our sample may not be bound to the host halo. However, consistent with previous surveys, we obtain a bound fraction (upper limit) of unity for halos with $M_{\rm halo} > 10^{11.5}$~\Msun\ out to $\approx2R_{\rm vir}$ \citep[]{Tumlinson_2011,Mathes_2014}. For the low mass halos with $M_{\rm halo} \leq 10^{11.5}$~\Msun, the bound fraction is $<$91\% within $R_{\rm vir}$ which declines to $<$79\% between $1-2 R_{\rm vir}$ and further declines to $<$43\% at $>2R_{\rm vir}$. We thus conclude that \OVI\ absorbers detected within $R_{\rm vir}$ are likely predominately bound to the host halos, even for dwarf galaxies. We verified that the results are consistent if we use $3 v_{\rm esc}(D)$ or $6 v_{\rm esc}(D)$ for galaxy-absorber association instead of a fixed LOS window of $\pm$300 \kms\ used here. A high bound fraction of $\lesssim 83$\% for \OVI\ absorbers in isolated dwarf galaxies (median \logm$\approx8.4$) at $D/R_{\rm vir}\lesssim2$ was recently observed in the CUBS survey \citep[]{Mishra_2024}. This is fully consistent with our findings (see Table~\ref{tab:OVI_bound_frac}).

Based on a small sample of 14 galaxies, \citet[]{Mathes_2014} reported that galaxies with halo mass $<10^{11.5}~\rm M_{\odot}$ have a bound fraction of only $<0.50$ in contrast to the $<0.91$ we measured here. This is likely because their \OVI\ measurements for low-mass galaxies are small in number and the majority of them have $D>R_{\rm vir}$ (7/8). The one \OVI\ component associated with a low-mass galaxy at $D<R_{\rm vir}$ is consistent with being bound to the host galaxy, in agreement with our findings. By comparing the \OVI\ kinematics with the radial velocities of a simulated galaxy ($M_{\rm halo}\approx 2\times10^{11}$~\Msun), the authors argued that the absorbing clouds with $|v_{\rm LOS}/v_{\rm esc}(D)|>1$ are likely to be outflowing winds since infalling gas does not exceed the escape velocity. Indeed, there is observational evidence that star-formation-driven outflows from dwarf galaxies can have velocities larger than the escape velocity \citep[see e.g.,][]{Romano_2023}. The $\approx20$\% of the \OVI\ components detected at $D> R_{\rm vir}$ in low-mass halos that exceed the escape velocities in our sample may indicate that they may stem from outflows. Alternatively, the 2-halo contributions, which become important at larger $D/R_{\rm vir}$,  can give rise to LOS velocities exceeding the escape velocity for low-mass galaxies \citep[see][]{Ho_2021}.

\subsection{Turbulence in the {\rm O~{\sc vi}}-bearing circumgalactic gas} 
\label{disc:turb}

Feedback due to AGN- and supernovae-driven winds and mergers can inject significant turbulence into the CGM, in addition to enriching it with metals. It is therefore important to obtain direct constraints on the turbulence in the CGM. In Sec.~\ref{sec:dop-b}, we reported that a substantial fraction of \OVI\ components require high turbulent velocity ($b_{\rm turb}\geq20$ \kms) to explain their observed $b$-parameters, assuming the maximum attainable gas temperature is equal to the virial temperature (see Fig.~\ref{fig:b-m_all}). Decomposing the $b$-parameters into thermal and turbulent components, and assuming that the thermal broadening is due to the virial temperature of the host halo, we obtain a range of $b_{\rm turb} \approx 5-60$~\kms\ with a median of 22~\kms\ and mean of 25 \kms.  \citet[]{Werk_2016} reported an average $b_{\rm turb}$ of $\approx40-50$~\kms\ for the \OVI\ absorbers in the COS-Halos sample, assuming the thermal broadening ranges from 6.4--16.2~\kms. Our inferred $b_{\rm turb}$ values are marginally lower than \citet{Werk_2016}. Recently, \citet{Chen_2023} reported $b_{\rm turb}$ of $\approx 5-30$~\kms\ for a sample of well-aligned \HI\ and low-ionization metal lines in the CGM of low-$z$ galaxies. \citet{Werk_2016} also inferred $b_{\rm turb} < 20$~\kms\ for the \SiIII\ (low-ionization) line. This may suggest that the different phases of the CGM are subject to different turbulent motions.

%%%%%%%%%%%%%%
\begin{figure*}
    \centering
    \includegraphics[width=0.5\linewidth]{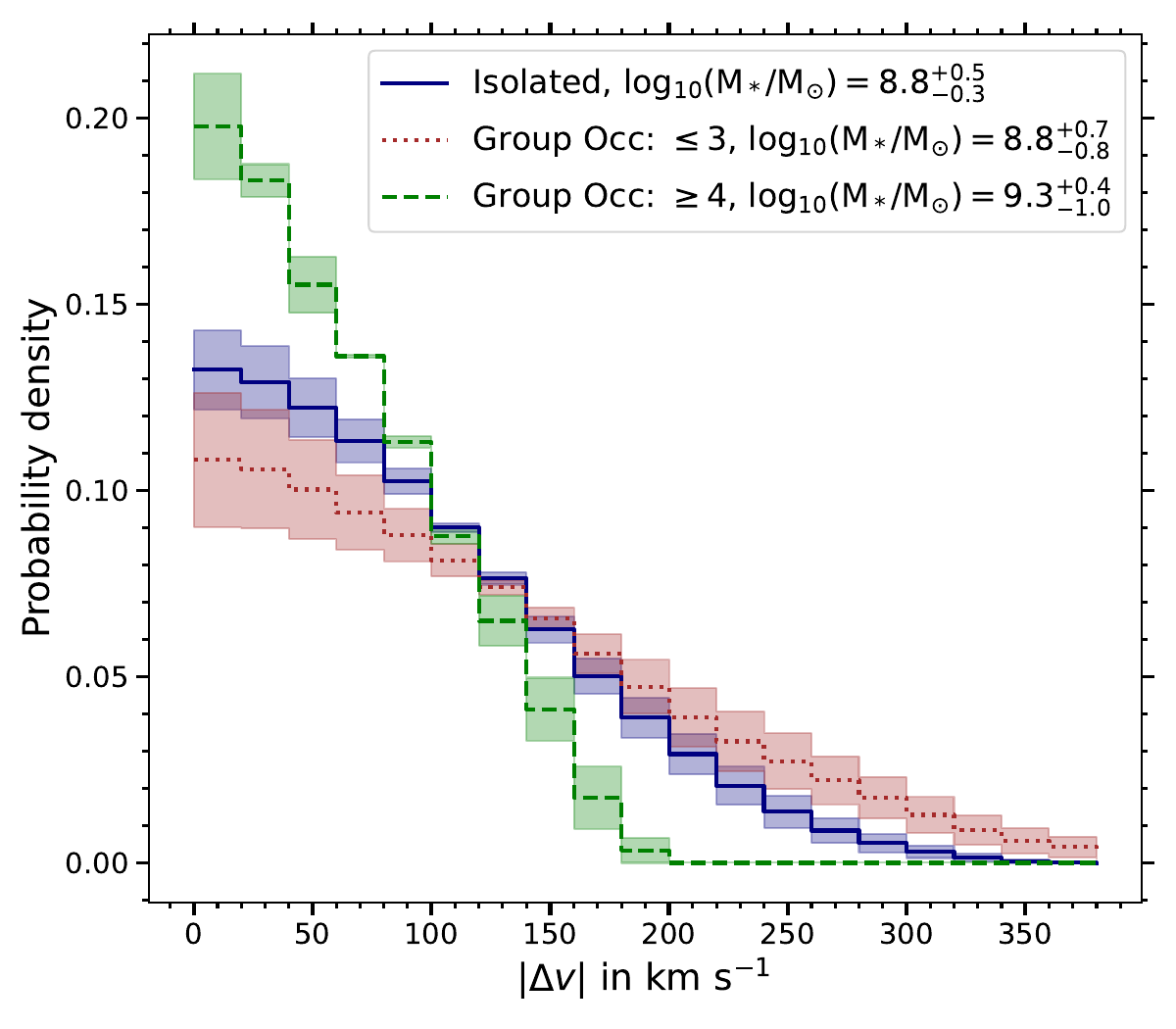}%
    \includegraphics[width=0.5\linewidth]{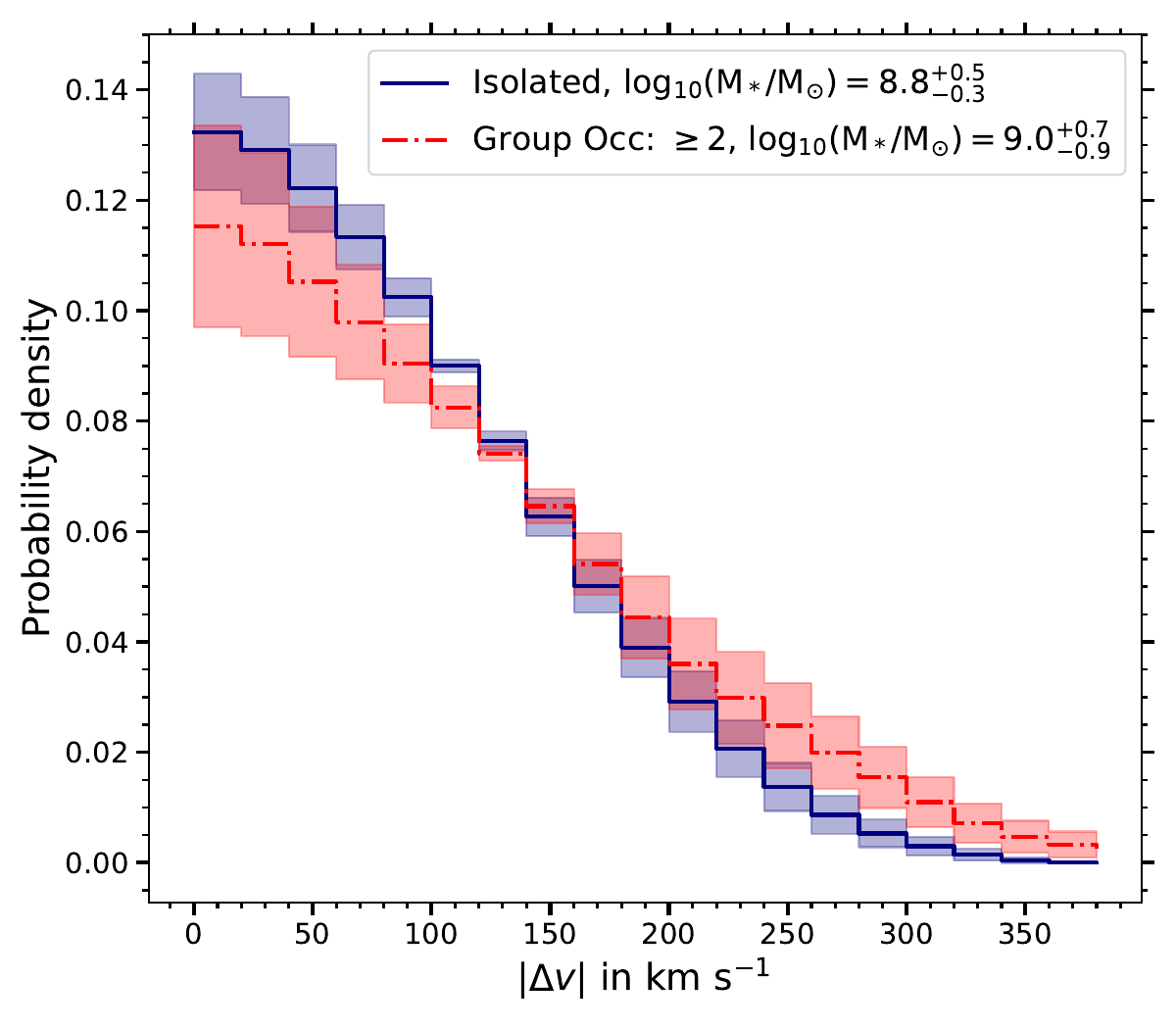}    
    \caption{{\tt Left:} TPCFs of \OVI\ absorption for isolated galaxies and group galaxies with occupancy, GOcc~$\leq3$ and $\geq4$ shown in solid blue, dotted brown, and dashed green lines, respectively. The shaded regions represent 68\% confidence intervals obtained with 100 bootstrap realizations. The stellar mass of the isolated sample is matched to the galaxies in the GOcc~$\leq 3$ group. The TPCF for the GOcc~$\geq 4$ group is significantly narrower compared to the rest. {\tt Right:} TPCF of \OVI\ absorption for the group galaxies following the definition of \citet[]{Pointon_2017} shown in red dash-dotted line. The TPCF for the isolated galaxies is the same as on the left. No significant difference is seen in the TPCFs of \OVI\ absorption for the two samples. The median stellar mass of the contributing galaxies is indicated in the legends, with subscripts and superscripts representing the 68\% range.} 
    \label{fig:tpcf_env_grp}
\end{figure*}
%%%%%%%%%%%%

Several galaxy-blind absorption line studies used aligned \HI--\OVI\ absorption pairs to solve for the thermal and non-thermal components of the $b$-parameter both at low $z$ \citep[e.g.,][]{Tripp_2008,Savage_2014} and high $z$ \citep[i.e.,][]{Muzahid_2012}. While the gas temperature turned out to be similar at low and high redshifts for the well-aligned \HI--\OVI\ pairs, the median $b_{\rm turb}$ was found to be 2--3 times larger at low-$z$ (i.e., 20--30~\kms). The median $b_{\rm turb}$ values obtained from blind absorption line studies are somewhat smaller than our inferred values but are consistent with the predictions of the theoretical models of \citet{Evoli_2011}.

Recently, based on both small-scale and large-scale simulations, \citet[]{Koplitz_2023} argued that $\sigma_{\rm LOS}$, defined as the standard deviation of velocity centroids of components along a sightline, is a better indicator of true line-of-sight turbulence than $b_{\rm turb}$, as it can capture the inter-cloud motions along with the random motions within individual clouds. Since most of our \OVI\ absorbers are comprised of one or two components, $\Delta v_{90}$ is a good proxy for the $\sigma_{\rm LOS}$. The moderate anti-correlation of $\Delta v_{90}$ with $D/R_{\rm vir}$ observed in this work (see Fig.~\ref{fig:v90_dn}) may indicate decreasing turbulence in the outer CGM. This is in agreement with the observed anticorrelation between velocity dispersion of \OVI\ absorbers and $D/R_{\rm vir}$ in \citet[]{Qu_2024}.  
% This is consistent with the recent findings of \citet[]{Qu_2024}. 
\citet[]{Borthakur_2016}, on the contrary, found an increasing trend in $\sigma_{\rm LOS}$, quantified as $b/\sqrt{2}$ of the strongest component, with $D/R_{\rm vir}$ based on their \HI\ measurements. We observe a moderate anti-correlation of $b$ with $D/R_{\rm vir}$ for our \OVI\ sample (see right panel of Fig.~\ref{fig:b-m_all}). We verified that selecting only the strongest component associated with the host galaxy does not change our results. Here we note that the galaxies in the \citet[]{Borthakur_2016} sample extend to a higher halo masses range (${\rm log}_{10}(M_{\rm halo}/\rm M_{\odot}) \approx 11.5 - 13$) as compared to our sample (${\rm log}_{10}(M_{\rm halo}/\rm M_{\odot}) \approx 10 - 12$). Additionally, different phases traced by \HI\ and \OVI\ may give rise to this apparent contrast in the turbulent velocity in the CGM.

\subsection{The connection between {\rm O~{\sc vi}} kinematics and galaxy environment}
\label{disc:kinematics_width}

In Section~\ref{sec:tpcf&v90} we showed that the overall kinematic extent of circumgalactic \OVI\ absorbers, characterized by the TPCF, is wider for high-mass galaxies. This is consistent with the findings of \citet{Ng_2019}. The difference becomes marginal ($2.3\sigma$) for the isolated subsample when the normalized \OVI\ TPCF is considered. However, this is not the case for the full (i.e., isolated plus group galaxies) sample (see Fig.~\ref{fig:tpcf_m_all}). The normalized \OVI\ TPCF for the low-mass subsample shows significantly higher probabilities at $|\Delta v| > v_{\rm vir}$ compared to the high-mass subsample.

The FoF algorithm used to identify isolated galaxies also provides us with a subsample of galaxies with one or more neighbors within $\pm500$~\kms\ of one another and a maximum projected distance of $\approx500$~kpc. We refer to this as our `group' subsample. The left panel of Fig.~\ref{fig:tpcf_env_grp} shows the TPCFs of \OVI\ absorption for the group galaxies with group-occupation number (GOcc) $\leq3$ and $\geq 4$ in dotted brown and dashed green lines, respectively. The shaded regions represent 68\% confidence intervals obtained with 100 bootstrap realizations. Note that we use an absorption system associated with a group only once in this exercise. The \OVI\ TPCF in solid blue line is for isolated galaxies with stellar mass matched to the galaxies in the GOcc~$\leq 3$ group (median \logm~$\approx8.8$). It is intriguing to note that the \OVI\ TPCF for the densest group (GOcc~$\geq4$), despite the higher median stellar mass of the member galaxies (median \logm~$\approx$~9.3), is significantly narrower compared to the other two subsamples. The difference between the mass-controlled isolated galaxies and GOcc~$\leq 3$ group galaxies is not that prominent. The GOcc~$\leq 3$ group shows a somewhat wider \OVI\ TPCF compared to the isolated subsample with a $4\sigma$ significance.

By comparing the \OVI\ TPCFs of isolated and group galaxies, \citet[]{Pointon_2017} reported that the velocity spread of \OVI\ in the CGM of group galaxies is significantly ($10\sigma$) narrower than that of isolated galaxies.\footnote{They defined a galaxy group as two or more galaxies with a LOS velocity difference of no more than 1000~\kms\ and located within 350 kpc (projected) of a background quasar sightline.} They suggested that the gas traced by \OVI\ in group galaxies is located at the interface of warm CGM of individual group galaxies, and the hotter region of the intra-group medium at the virial temperature. The narrow profiles are, according to them, a consequence of low-column density gas in the higher velocity wings being further ionized due to higher temperature. 
We see such a stark difference only for the GOcc~$\geq4$ group sample. Out of the 6 groups that show detectable \OVI\ absorption in the sample of \citet{Pointon_2017}, only one has GOcc~$\geq 4$. If we adopt a group definition similar to \citet[]{Pointon_2017}, we find {\em a marginal $2\sigma$} difference between the \OVI\ TPCFs of isolated and group subsamples, with a somewhat broader kinematic profile for the group galaxies (see right panel of Fig.~\ref{fig:tpcf_env_grp}). Since we observed a strong mass dependence for the \OVI\ TPCF (lower-left panel of Fig.~\ref{fig:tpcf_m_all}), this marginal difference can be attributed to the slightly higher median mass for the group galaxies. It is therefore important to control for the stellar mass, particularly when the \OVI\ TPCF is not normalized by the circular velocity.

It is worthwhile to note that a significant fraction of groups with large occupation numbers (GOcc~$\geq4$) do not show any detectable \OVI\ absorption. Out of the six groups with GOcc~$\geq4$, only two show detectable \OVI\ absorption, totaling 3 Voigt profile components with $b$-parameters of $11.2\pm2.9$, $15.0\pm2.4$, and $36.7\pm4.9$~\kms. None of these components are consistent with temperatures $\geq 10^{5.5}$~K when compared\footnote{We decompose the $b$-parameter into thermal and non-thermal components using the $b(\OVI)$ and $b(\HI)$.} with the $b$ value of the nearest \HI\ component. In fact, the inferred gas temperature for these three components (i.e., a few~$\times 10^{4}$~K) is similar to what is expected from photoionized gas. Evidently, the narrowness of the \OVI\ TPCF for the GOcc~$\geq 4$ group is not because of the gas being overheated. Nevertheless, the non-detection of \OVI\ for a significant fraction of groups with GOcc~$\geq4$  \citep[i.e., $\approx67$\% our sample and $\approx50$\% for][]{Pointon_2017} is intriguing. This could be an indication that the medium is too hot for \OVI.

To further investigate the non-detection of \OVI\ in 4/6 GOcc~$\geq4$ groups in our sample, we found that in every group there is at least one member galaxy with an impact parameter $<R_{\rm vir}$. In fact, there are a total of 11 member galaxies in those 4 groups with $D < R_{\rm vir}$. However, only 2/11  galaxies are classified as `SF'. Thus, the lack of \OVI\ in these groups can also be attributed to the lower \OVI\ covering fraction in the CGM of passive galaxies, as shown in Paper I  \citep[see also][]{Oppenheimer_2016, Tchernyshyov_2022}.

\section{Summary}
\label{sec:summary}

In this paper, we present a study of the kinematics of \OVI-bearing gas around 60 low-redshift ($0.1\lesssim z\lesssim0.7$) galaxies with a median [68\% range] \logm~$=8.9$ [8.1--9.7] and median [68\% range] $D=115$~kpc [64--192 kpc] observed in 16 quasar fields in the MUSEQuBES survey. The \OVI\ absorption systems detected around these galaxies enabled us to investigate the connections between galaxy properties on the \OVI\ kinematics. Our key findings are:

\begin{itemize}

    \item We found that $\approx 83$\% of all the \OVI\ components detected within $\pm300$~\kms\ of the galaxy redshifts have $v_{\rm LOS}/v_{\rm esc}(D) < 1$. The fraction increases to 94\% when only galaxies with $D<R_{\rm vir}$ are considered. Even for dwarf galaxies (\logm~$<9$),  $\approx91$\% of the \OVI\ components detected within $R_{\rm vir}$ are consistent with being bound to the host halo (Table \ref{tab:OVI_bound_frac}).

    \item The $b$-parameter of the \OVI\ components and the $\Delta v_{90}$ of the \OVI\ systems associated with the galaxies show a mild anti-correlation with $D/R_{\rm vir}$. Significantly larger $b$ and $\Delta v_{90}$ values, compared to galaxy-blind \OVI\ absorbers, are seen at $D\lesssim1.5 R_{\rm vir}$ (Figs.~\ref{fig:b-m_all} \& ~\ref{fig:v90_dn}).

    \item A positive trend is observed between the $b$-parameter and $M_{\star}$ for the isolated galaxies ($\tau = 0.25, p = 0.02$; Fig.~\ref{fig:b-m_all}). Assuming that the thermal broadening is due to the virial temperature of the host halo, we determine non-thermal broadening to be in the range 5--60~\kms, with a median of 22~\kms.

    \item The velocity TPCF of \OVI\ absorption for the high-mass galaxies (\logm~$>9$) is significantly broader compared to their low-mass counterpart. However, the difference becomes marginal for the isolated subsample when the TPCF is normalized by the circular velocity  (Fig.~\ref{fig:tpcf_m_all}). We do not find any significant difference between the TPCF of isolated and group (with GOcc~$\geq2$) subsamples when the stellar mass is matched  (Fig.~\ref{fig:tpcf_env_grp}; right).

    \item The \OVI\ TPCF of the group subsample with GOcc~$\geq4$ is significantly narrower compared to the isolated subsample and group subsample with GOcc~$\leq3$ (Fig.~\ref{fig:tpcf_env_grp}; left). The observed $b$-parameters of the two \OVI\ systems contributing to the TPCF indicate that the narrowness of the TPCF is {\em not} due to high gas temperatures. However, we note that 4/6 groups with GOcc~$\geq4$ do not exhibit any detectable \OVI\ absorption which may be an indication that the medium is too hot for \OVI. The prevalence of passive galaxies in such galaxy overdensity can also explain the lack of \OVI\ detection. As such, we do not find any convincing evidence that the \OVI\ kinematics depend on the galaxy environment.

\end{itemize}

\begin{acknowledgments}
 SD and SM acknowledge support from the Indo-Italian Executive Programme
of Scientific and Technological Cooperation 2022—2024 (TPN:
63673). SD acknowledges Prof. R. Srianand, Dr. Aseem Paranjape for insightful discussions.
\end{acknowledgments}

\vspace{5mm}

\bibliography{all_ref_ovi_apj}{}
\bibliographystyle{aasjournal}

% \bibliographystyle{mnras}
% \bibliography{all_ref_ovi}



\end{document}